\documentclass[a4paper,10pt]{article}
\usepackage[utf8]{inputenc}
\usepackage{authblk}

\usepackage{graphicx}
\usepackage{epstopdf}
\usepackage{todonotes}
\usepackage{lineno,hyperref}
\usepackage{upgreek}
\usepackage{units}
\usepackage{multirow}
\usepackage{amssymb}	
\usepackage{amsmath}	

\usepackage{caption}
\usepackage{subcaption}

\title{Characterisation of strip silicon detectors for the ATLAS Phase-II Upgrade with a micro-focused X-ray beam}

\author[1]{Luise Poley}
\author[2]{Andrew Blue}
\author[2]{Richard Bates}
\author[1]{Ingo Bloch}
\author[1]{Sergio D\'{i}ez}
\author[3]{Javier Fernandez-Tejero}
\author[3]{Celeste Fleta}
\author[4]{Bruce Gallop}
\author[5]{Ashley Greenall}
\author[1]{Ingrid-Maria Gregor}
\author[6]{Kazuhiko Hara}
\author[7]{Yoichi Ikegami}
\author[8]{Carlos Lacasta}
\author[1]{Kristin Lohwasser}
\author[2]{Dzmitry Maneuski}
\author[9]{Sebastian Nagorski}
\author[10]{Ian Pape}
\author[4]{Peter W. Phillips}
\author[11]{Dennis Sperlich}
\author[10]{Kawal Sawhney}
\author[8]{Urmila Soldevila}
\author[3]{Miguel Ullan}
\author[7]{Yoshinobu Unno}
\author[12]{Matt Warren}

\affil[1]{Deutsches Elektronen-Synchrotron, Notkestra{\ss}e, Hamburg, Germany}
\affil[2]{SUPA School of Physics and Astronomy, University of Glasgow, University Avenue, Glasgow, United Kingdom}
\affil[3]{Centro Nacional de Microelectr\'{o}nica (IMB-CNM), Consejo Superior de Investigaciones Cient\'{i}ficas, Campus UAB-Bellaterra, Barcelona, Spain}
\affil[4]{Particle Physics Department, STFC Rutherford Appleton Laboratory, Harwell Science and Innovation Campus, Didcot, United Kingdom}
\affil[5]{Department of Physics, University of Liverpool, Cambridge Street, Liverpool, United Kingdom}
\affil[6]{Institute of Pure and Applied Sciences, University of Tsukuba, Tennodai, Tsukuba, Japan}
\affil[7]{Institute of Particle and Nuclear Study, KEK, Oho, Tsukuba, Japan}
\affil[8]{IFIC, CSIC-U. Valencia, c/ Catedr\'{a}tico Jos\'{e} Beltr\'{a}n, Paterna, Spain}
\affil[9]{School of Physics, Dublin Institute of Technology, Kevin Street, Dublin, Ireland}
\affil[10]{Diamond Light Source Ltd, Diamond House, Harwell Science and Innovation Campus, Didcot, United Kingdom}
\affil[11]{Institut für Physik, Humboldt-Universit\"{a}t zu Berlin, Newtonstra{\ss}e, Berlin, Germany}
\affil[12]{Department of Physics and Astronomy, University College London, Gower Street, London, United Kingdom}

\begin{document}

\maketitle

\begin{abstract}
The planned HL-LHC (High Luminosity LHC) in 2025 is being designed to maximise the physics potential through a sizable increase in the luminosity up to \unit[$6\cdot10^{34}$]{$\text{cm}^{-2}\text{s}^{-1}$}. A consequence of this increased luminosity is the expected radiation damage at \unit[3000]{$\text{fb}^{-1}$} after ten years of operation, requiring the tracking detectors to withstand fluences to over $1\cdot10^{16}$ \unit[1]{MeV n$_{\text{eq}}$/cm$^2$}. In order to cope with the consequent increased readout rates, a complete re-design of the current ATLAS Inner Detector (ID) is being developed as the Inner Tracker (ITk).

Two proposed detectors for the ATLAS strip tracker region of the ITk were characterized at the Diamond Light Source with a \unit[3]{$\upmu$m} FWHM \unit[15]{keV} micro focused X-ray beam. The devices under test were a \unit[320]{$\upmu$m} thick silicon stereo (Barrel) ATLAS12 strip mini sensor wire bonded to a \unit[130]{nm} CMOS binary readout chip (ABC130) and a \unit[320]{$\upmu$m} thick full size radial (end-cap) strip sensor - utilizing bi-metal readout layers - wire bonded to \unit[250]{nm} CMOS binary readout chips (ABCN-25).

A resolution better than the inter strip pitch of the \unit[74.5]{$\upmu$m} strips was achieved for both detectors. The effect of the p-stop diffusion layers between strips was investigated in detail for the wire bond pad regions.

Inter strip charge collection measurements indicate that the effective width of the strip on the silicon sensors is determined by p-stop regions between the strips rather than the strip pitch.
\end{abstract}

\section{Introduction}

Around 2025, the Large Hadron Collider (LHC) at CERN will be upgraded to an instantaneous luminosity of 
$\mathcal{L} = \unit[6\cdot10^{34}]{\text{cm}^{-2}\text{s}^{-1}}$ from the design luminosity of $\mathcal{L} = \unit[1\cdot10^{34}]{\text{cm}^{-2}\text{s}^{-1}}$ (High Luminosity LHC)~\cite{1}. The corresponding upgrade of the ATLAS detector at the LHC will require the replacement of the current ID~\cite{ID} with a new Inner Tracker (ITk)~\cite{2}.

Unlike the current Inner Detector (ID), the ITk will be an all-silicon tracker, constructed to maintain tracking performance in the high occupancy environment and to cope with the increase of approximately a factor of ten in the total radiation fluence. New technologies are used to ensure that the system can survive this harsh radiation environment and to optimise the distribution of material. A new readout scheme allows the implementation of a track trigger, which will significantly improve the ATLAS data taking capabilities.

The ITk will consist of both pixel and strip detectors, consisting of a central barrel region (with sensor strips are aligned parallel to the beam) between \unit[$\pm\,1.4$]{m} and two end-caps (with sensor strips orthogonal to the beam) that extend the length of the strip detector to \unit[$\pm\,3$]{m}. The detectors cover \unit[$\pm\,2.5$]{units of pseudorapidity}~\cite{2}.

Silicon strip sensors for the future ATLAS detector have been developed, and detector module prototypes have been constructed~\cite{3}. In ongoing R\&D efforts, current versions of module components are being improved towards designs suitable for the future ATLAS strip tracker. 

This paper describes the use of a micro focused X-ray beam for precise scans across several sensor strips~\cite{4,5}. The scan was conducted to compare the signals collected for two silicon strip detector modules from two successive module designs. As well as investigating details of the new ATLAS12 strip layout, this also allowed the performance of the binary readout chip (ABC130) to be compared to its predecessor (ABCN-25).

\section{Devices}

The devices investigated in the testbeam were constructed as similarly as possible to the required future strip tracker modules, given current availability of components~\cite{6}. Each module consisted of a prototype of an actual silicon micro strip sensor, onto which a hybrid with front-end readout chips was glued~\cite{7}. Aluminium wedge wire bonding was used both to connect the sensor strips and ASIC readout channels and to connect the ASICs and hybrid electrically. Both tested devices are described in detail in the following sections~\ref{ss:d25} and~\ref{ss:d13}. The devices were mounted on test frames, which provided high voltage to bias the sensor as well as low voltage to power the hybrid and an interface to the readout electronics.

The strip sensors used in both modules were AC-coupled with n-type implants in a p-type float-zone silicon bulk (n in p FZ). Signals were read out using an FPGA development board (ATLYS) with the C++ based framework SCTDAQ, which was developed for the Semiconductor Tracker (SCT) in the current ATLAS detector and modified for the operation of modules for the future ATLAS strip tracker.

In preparation for test beam operation, both devices were tested in the lab using the SCTDAQ framework. The performance of all readout channels, such as input noise and gain, was determined using known injected charges, generated on calibration capacitors present on the ASIC. Using these measurements, all channels were trimmed to have comparable thresholds: in absence of any signal in the sensor, all channels were adjusted to use the same threshold starting point (corresponding to \unit[0]{fC} input charge), see table~\ref{tab:scanpar}.

\subsection{ABCN-25 end-cap module}
\label{ss:d25}

\begin{figure}
\centering
\includegraphics[width=\textwidth]{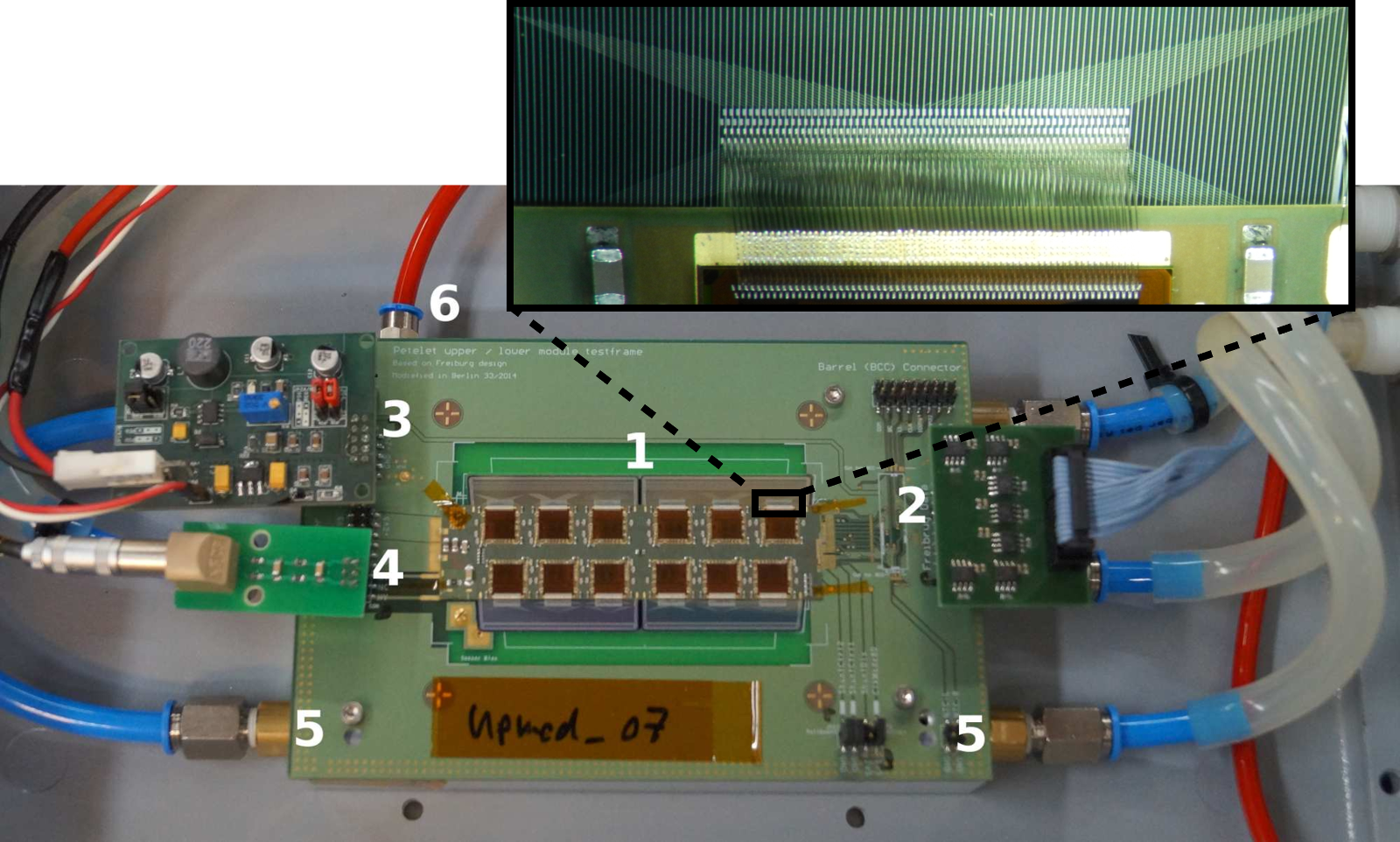}
\caption{ABCN-25 end-cap module as operated in the test beam. The module (1), consisting of two sensors, connected by a hybrid with twelve readout chips, was mounted on a test frame with a connector for data readout (2), low voltage (3) to power the module, high voltage (4) to bias the sensor, cooling tubes (5) leading through the mounting jig and a vacuum line (6) for a good contact between module and cooling jig. The sensor strip orientation is shown in a magnified sensor area with wire bond connections between strips and readout channels.}
\label{fig:device_e}
\end{figure}

The module built using ABCN-25 binary readout chips was an end-cap module. It consisted of a hybrid with twelve ABCN-25 readout chips~\cite{8} that had been glued to two silicon sensors (see Figure~\ref{fig:device_e}). 
The silicon is \unit[320]{$\upmu$m} thick and each sensor comprises two rows of parallel strips with a combined length of \unit[38.7]{mm}. The circular shape of an end-cap leads to wedge shaped sensors with strip pitches increasing towards outer radii from \unit[93]{$\upmu$m} to \unit[106]{$\upmu$m}~\cite{9}.

The module was operated at a reverse bias voltage of \unit[120]{V} (over-depleted as the voltage for full depletion of the p-n diode is $V_{\text{depletion}} \approx\unit[50]{\text{V}}$). The device was operated on an aluminium plated jig cooled down to \unit[10]{$^{\circ}$C} by a chiller to compensate for the heat produced by the readout chips (\unit[$\approx 300$]{mW} per ABCN-25 ASIC)~\cite{8}.

\subsection{ABC130 Mini Module}
\label{ss:d13}

The baseline design of the new strips tracker envisages binary readout by means of the ABC130 readout ASIC~\cite{7, 12}, fabricated in the 8RF \unit[130]{nm} CMOS technology from IBM Semiconductors. The first batch of ABC130 was produced and delivered in November 2013 and is currently under test. The current proposal is that any future CMOS ASIC for the ITK Strips will utilise the same front-end designs.

Preliminary tests of the initial production of the ABC130 chips have been performed. Communication and configuration of the chips work correctly. Data readout at a rate of \unit[80]{MHz} has been achieved, and the current consumption of the chip has been reduced significantly (as expected as a consequence of the \unit[130]{nm} process) to give an estimated \unit[3]{W/module}, as compared to the current \unit[20]{W/module} for the present ABCN-25 chip fabricated in a \unit[250]{nm} process.

\begin{figure}
\centering
\includegraphics[width=0.7\textwidth]{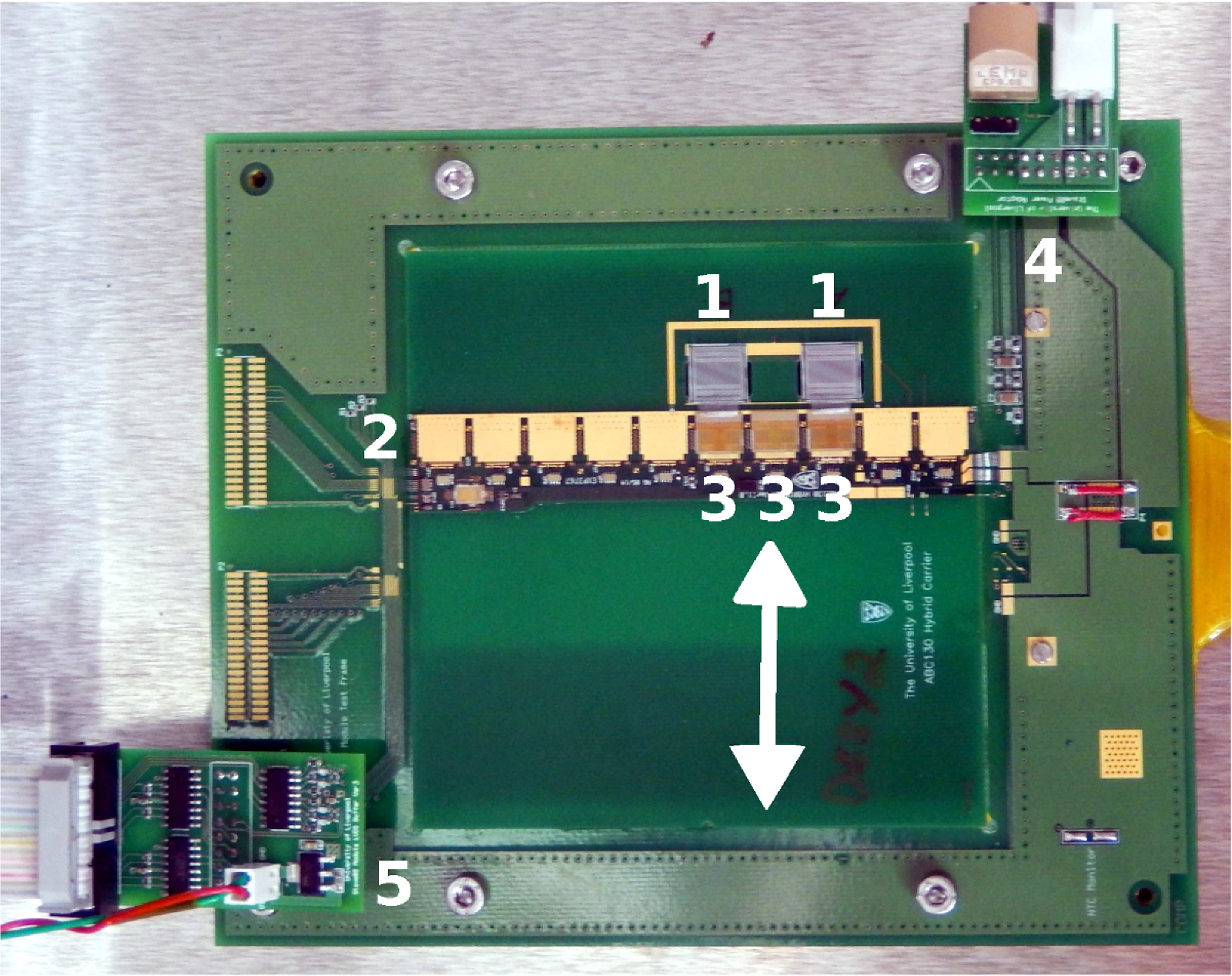}
\caption{ABC130 mini module mounted on a test frame. The device consists of two miniature prototypes of ATLAS silicon strip sensors (1), a hybrid (2) with three ABC130 readout chips (3), wire bonded to the miniature sensors. The device is mounted on a test frame, providing power (4) and an interface to the readout electronics (5). The arrow indicates the sensor strip orientation.}
\label{fig:device_b}
\end{figure}

To further test the new ABC130 readout chip and ATLAS12 strip layout, a mini version of a full-scale module was assembled. Three ABC130 chips were glued to an FR4 hybrid, with two of these wire bonded to ATLAS12 mini sensors~\cite{13}, see figure~\ref{fig:device_b}.
The mini sensors under test consisted of 104 strips (\unit[74.5]{$\upmu$m} pitch in 2 rows) with punch through protection (PTP) structures and an overall size of \unit[$1\times1$]{$\text{cm}^2$}. 

All data were multiplexed through the hybrid control chip (HCC), and routed via a custom designed PCB along with HV and LV connections. The HCC interfaces the ABC130 ASICs on the hybrid to the to end of structure electronics.

A reverse bias of \unit[300]{V} was applied to fully deplete the mini sensors. The low power output meant that the detector did not have to be cooled whilst in operation.

\section{Sensor scans at the Diamond Synchrotron}

\subsection{X-ray Beam}

The measurements presented here were all performed at the B16 beamline at the Diamond Synchrotron Light Source. This beam line comprises of a water-cooled fixed-exit double crystal monochromator that is capable of providing monochromatic beams over a \unit[$4-20$]{keV} photon energy range. An unfocused monochromatic beam is provided to the experimental hutch. A compound refractive lens (CRL) was used to produce a \unit[15]{keV} micro-focused X-ray beam.

\begin{figure}
\centering
\begin{subfigure}{.5\textwidth}
  \centering
  \includegraphics[width=\linewidth]{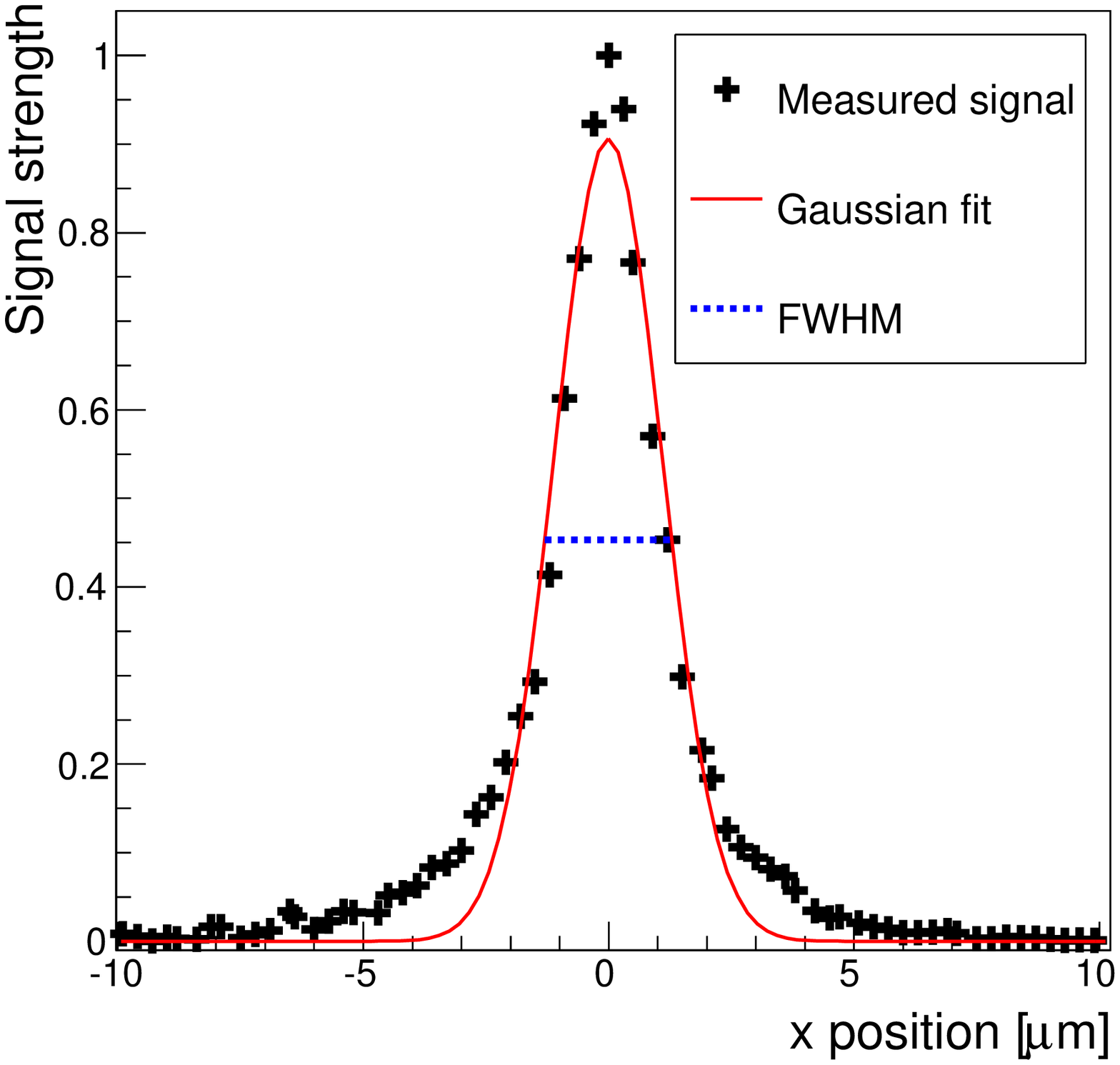}
  \caption{Beam width in x: {\unit[2.6]{$\upmu$m}}}
  \label{fig:sub1}
\end{subfigure}%
\begin{subfigure}{.5\textwidth}
  \centering
  \includegraphics[width=\linewidth]{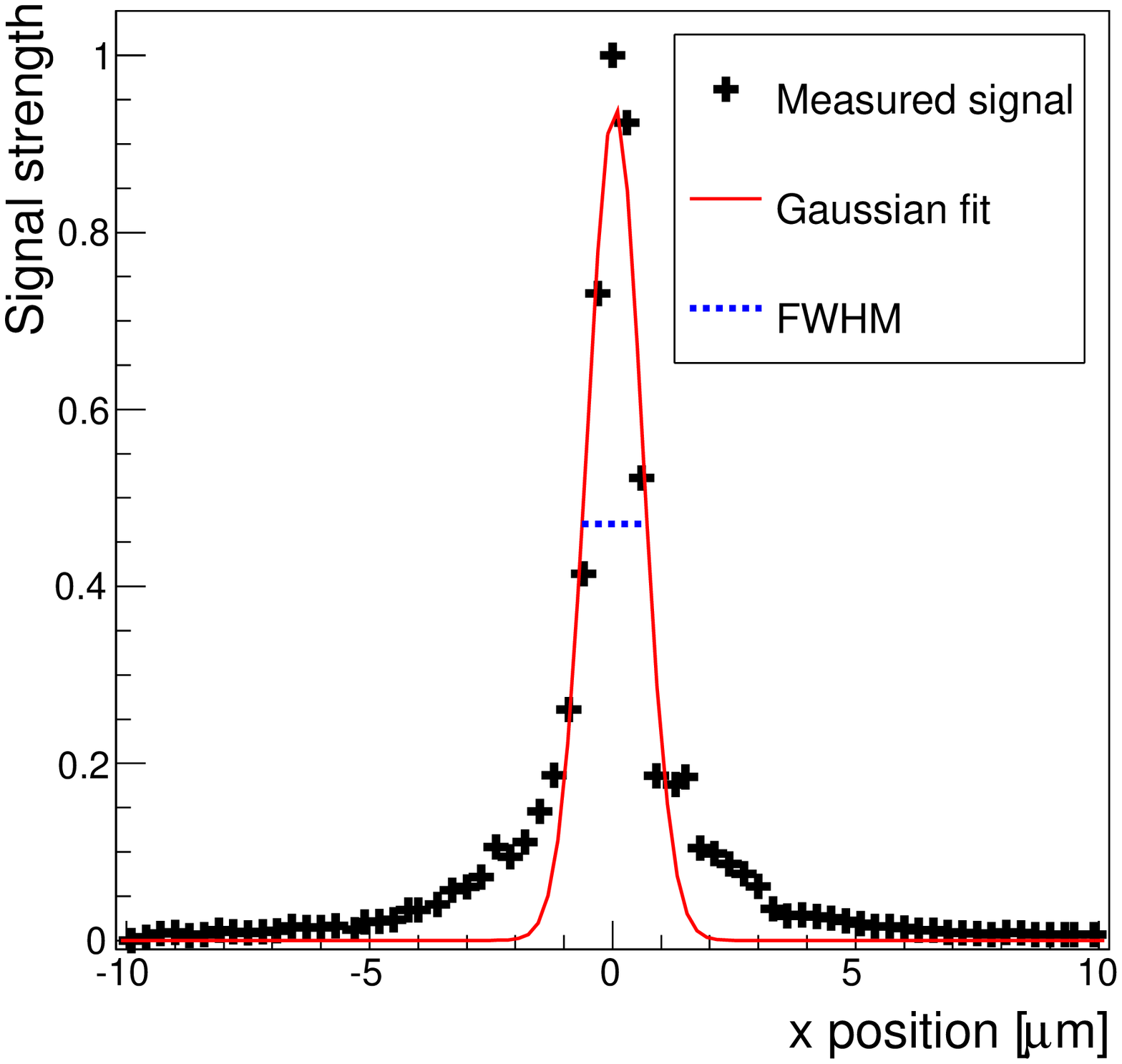}
  \caption{Beam width in y: {\unit[1.3]{$\upmu$m}}}
  \label{fig:sub2}
\end{subfigure}
\caption{Diamond X-ray beam profiles in x and y measured using a gold wire.}
\label{fig:test}
\end{figure}

The size of the micro-focused beam was determined by measuring transmissions scans with a \unit[200]{$\upmu$m} gold wire. Scans were made in both x and y across the beam, and the derivative of these scans indicated the beam size to have a sigma of \unit[2.6]{$\upmu$m} and \unit[1.3]{$\upmu$m} in the vertical and horizontal directions respectively (see figures~\ref{fig:sub1} and~\ref{fig:sub2}).

\subsection{DAQ and readout}

For data acquisition, a machine trigger from the beam was reduced from \unit[2]{MHz} to \unit[1]{kHz} and fed in to the Digilent ATLYS readout board utilising a Xilinx Spartan 6 LX45 FPGA. The devices were located on a stage which could be moved precisely, via custom Python scripts, in 3 dimensions (x, y and z). A signal was sent after the stage movement was complete to the DAQ to start the data acquisition. This allowed for a highly automated, fast and efficient run control for all scans taken during the allocated beam time.

\begin{figure}
\centering
\begin{subfigure}{.49\textwidth}
  \centering
  \includegraphics[width=\linewidth]{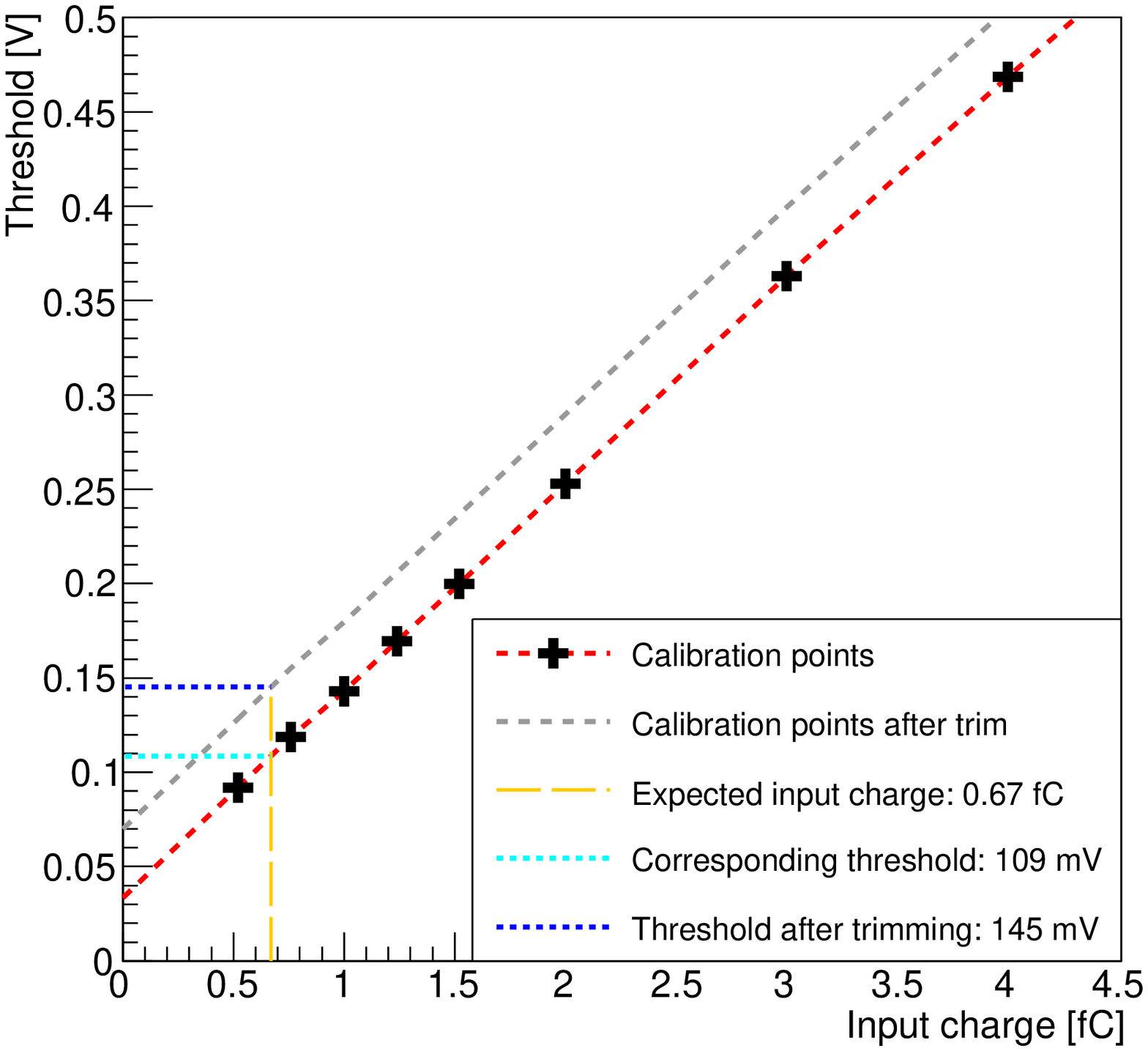}
  \caption{Calibration curve for ABCN-25 device: the threshold corresponding to the expected input charge is {\unit[109/145]{mV}} before/after trimming.}
  \label{fig:cal1}
\end{subfigure}
\begin{subfigure}{.01\textwidth}
\hfill
\end{subfigure}%
\begin{subfigure}{.49\textwidth}
  \centering
  \includegraphics[width=\linewidth]{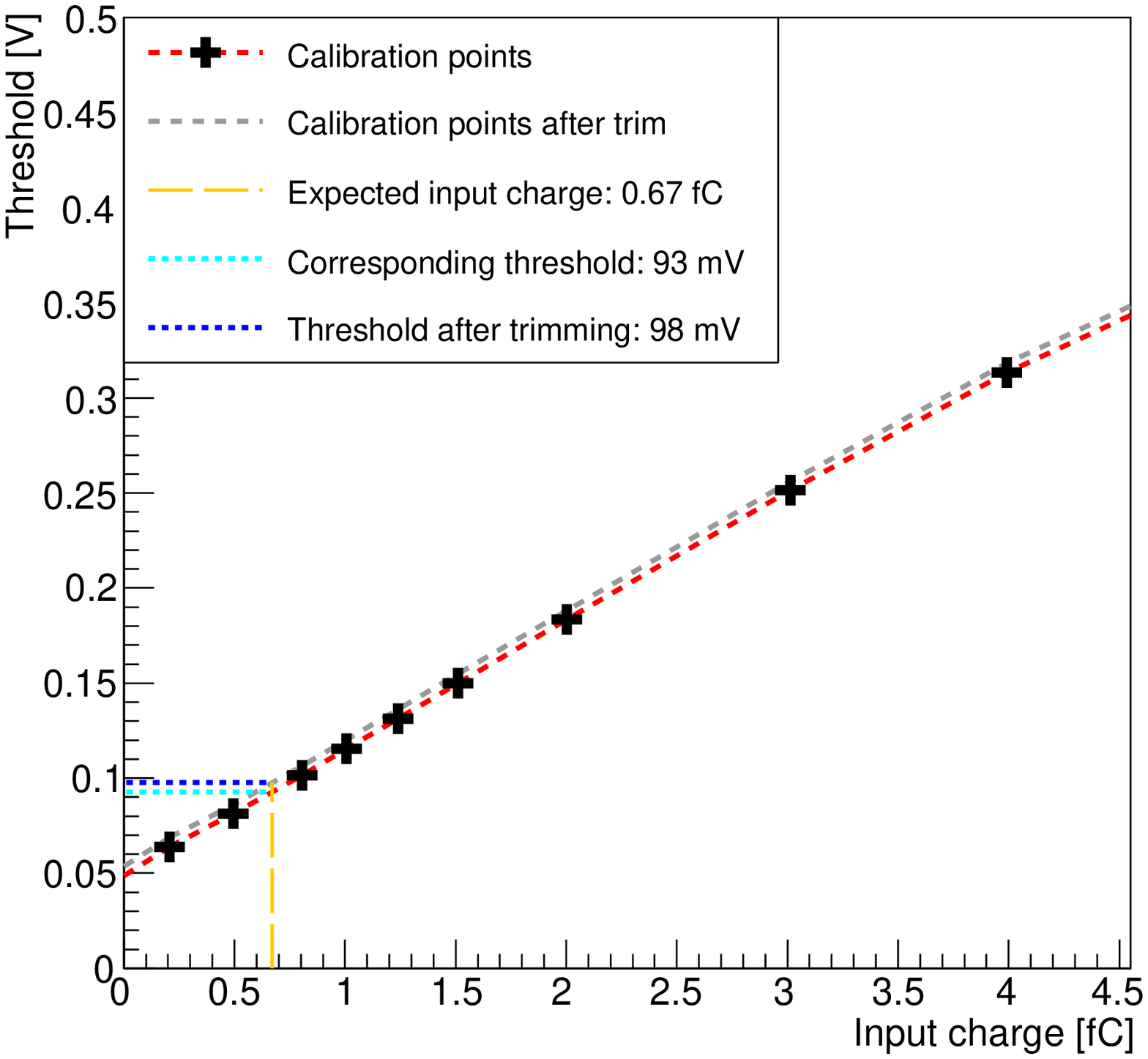}
  \caption{Calibration curve for ABC130 device: the threshold corresponding to the expected input charge is {\unit[93/98]{mV}} before/after trimming.}
  \label{fig:cal2}
\end{subfigure}
\caption{Threshold settings corresponding to different input charges, determined by internal calibration circuits of the devices under test (red lines). Due to internal noise of the ASIC, an input charge of {\unit[0]{fC}} corresponds to threshold levels of {\unit[30/40]{mV}} for the ABCN-25/ABC130 device (without trimming). Each data point (black markers) represents an average of the input channels of one ASIC. Channels with higher thresholds were used as an input to raise the thresholds of all channels to the same level (dashed grey line) for uniformity over a whole ASIC, leading to overall higher threshold levels. As a consequence, the threshold corresponding to the expected input charge for single photon events (orange line) increases (dark blue line compared to light blue line).}
\label{fig:cal}
\end{figure}

With both detectors operating a \unit[40]{MHz} clock, it was expected based on the beam current that there was average of 2-3 photons traversing the silicon sensors in the \unit[25]{ns} collection time. Since each \unit[15]{keV} photon has \unit[51.07]{\%} chance of interaction with \unit[320]{$\upmu$m} silicon~\cite{14}, the majority of acquisitions contained single X-ray photon events, with each \unit[15]{keV} interaction depositing about 4200 electrons (due to the \unit[3.6]{eV} electron hole pair creation energy of silicon). This is equivalent to \unit[0.67]{fC}, corresponding to a threshold of about \unit[100]{mV} (trimming: $\unit[0]{\text{fC}}\,\widehat{=}\,\unit[54]{\text{mV}}$) in the ABC130 device and about \unit[150]{mV} (trimming: $\unit[0]{\text{fC}}\,\widehat{=}\,\unit[70]{\text{mV}}$) in the ABCN-25 device (see figures~\ref{fig:cal1} and~\ref{fig:cal2}).
All data were written to disk by the DAQ as custom histograms, and analysis scripts written in ROOT to study signal responses at various positions and thresholds.

\section{Performed scans}
\label{s:pf}

\begin{figure}
\centering
\includegraphics[width=0.45\textwidth]{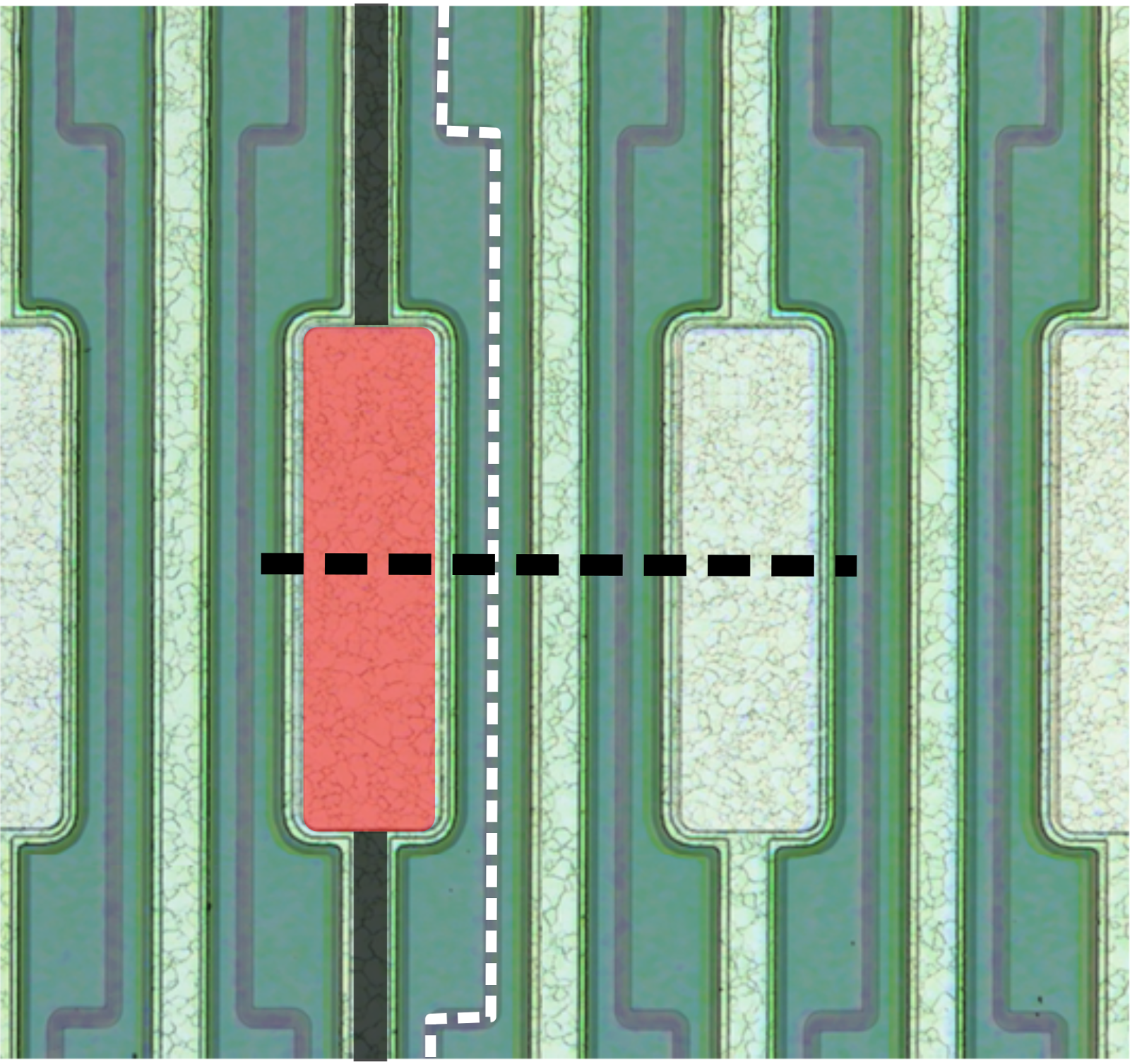}
\caption{Silicon strip sensor with a strip pitch of {\unit[74.5]{$\upmu$m}}. Aluminium bond pads (red area) are added to the strips (dark grey area) to allow for electrical connections between sensor strips and readout channels by wire bonds. With the width of the bond pads being close to the strip pitch, the p-stops (dashed white line) need to be guided around the bond pads, leading to a larger distance to the strip implant beneath the bond pad and a smaller distance to the adjacent strip implant without a bond pad. The dashed black line indicates the approximate beam positions during the scan.}
\label{fig:pstops}
\end{figure}

The X-ray beam profile, \unit[$2.6 \cdot 1.3$]{$\upmu$m$^2$}, was sufficiently small compared to the strip pitch to resolve the transverse sensor structure by performing threshold scans for different beam positions across the strips. The two devices (end-cap sensor with ABCN-25 hybrid and barrel sensor with ABC130 hybrid) were positioned differently in the X-ray beam: for the end-cap sensor the stage scan caused the beam to traverse a set of strips some distance from the bond pads while for the barrel sensor the beam traversed the region of the bond pads. 
Threshold scans across three sensor strips were performed for both devices: for each position of the beam on the sensor, the same number of triggers was sent to the readout system for a range of thresholds in order to investigate the input charge in different regions between sensor strips with a binary readout system.
Table~\ref{tab:scanpar} shows an overview of the scanning parameters for both devices at each position of the stage.
\begin{table}[htp]
\centering
\begin{tabular}{l|cc}
 & End-cap module & Barrel module\\
\hline
Strip pitch, $[\upmu$m$]$ & 103 & 74.5\\
p-stop & regular & irregular\\
Scan length, $[\upmu$m$]$ & 190 & 210 \\
Position step size, $[\upmu$m$]$ & 10 & 5 \\
Threshold range, $[$mV$]$ & 80.0-444.8 & 73.2-397.8\\
Trimming & \unit[0]{fC} = \unit[70]{mV} & \unit[0]{fC} = \unit[53.6]{mV} \\
Triggers per threshold & 3500 & 10000 \\
Readout chips & ABCN-25 & ABC130 \\
\end{tabular}
\caption{Overview of the scanning parameters performed for two silicon strip sensors}
\label{tab:scanpar}
\end{table}
With these scans, the influence of p-stop regions on the overall signal shape was also investigated. P-stops are implanted in the p-doped sensor between the n-doped strips in order to avoid short-circuits on the sensor surface after irradiation. On most of the sensor area, p-stops are implanted parallel to the strips with equal distances to both adjacent strips, but where the strips are connected to the ASIC readout channels via wire bonds, p-stops are placed around the required aluminium bond pads (see figure~\ref{fig:pstops}). 
Positioning p-stops at a sufficient distance from the wire bond pads leads to uneven distances to the adjacent sensor strips on the left and right side of the p-stop. In the ATLAS12 mini-sensors under test, the wire bond pad structures account for one third of the sensor surface.

The effects of these p-stop shapes were investigated by scanning across one of these regions of the sensor and are shown in section~\ref{ss:esw}. Comparative scans were also made in the centre of the conventional p-stop doping regions.

\section{Results}

\begin{figure}
\centering
\begin{subfigure}{.49\textwidth}
  \centering
  \includegraphics[width=\linewidth]{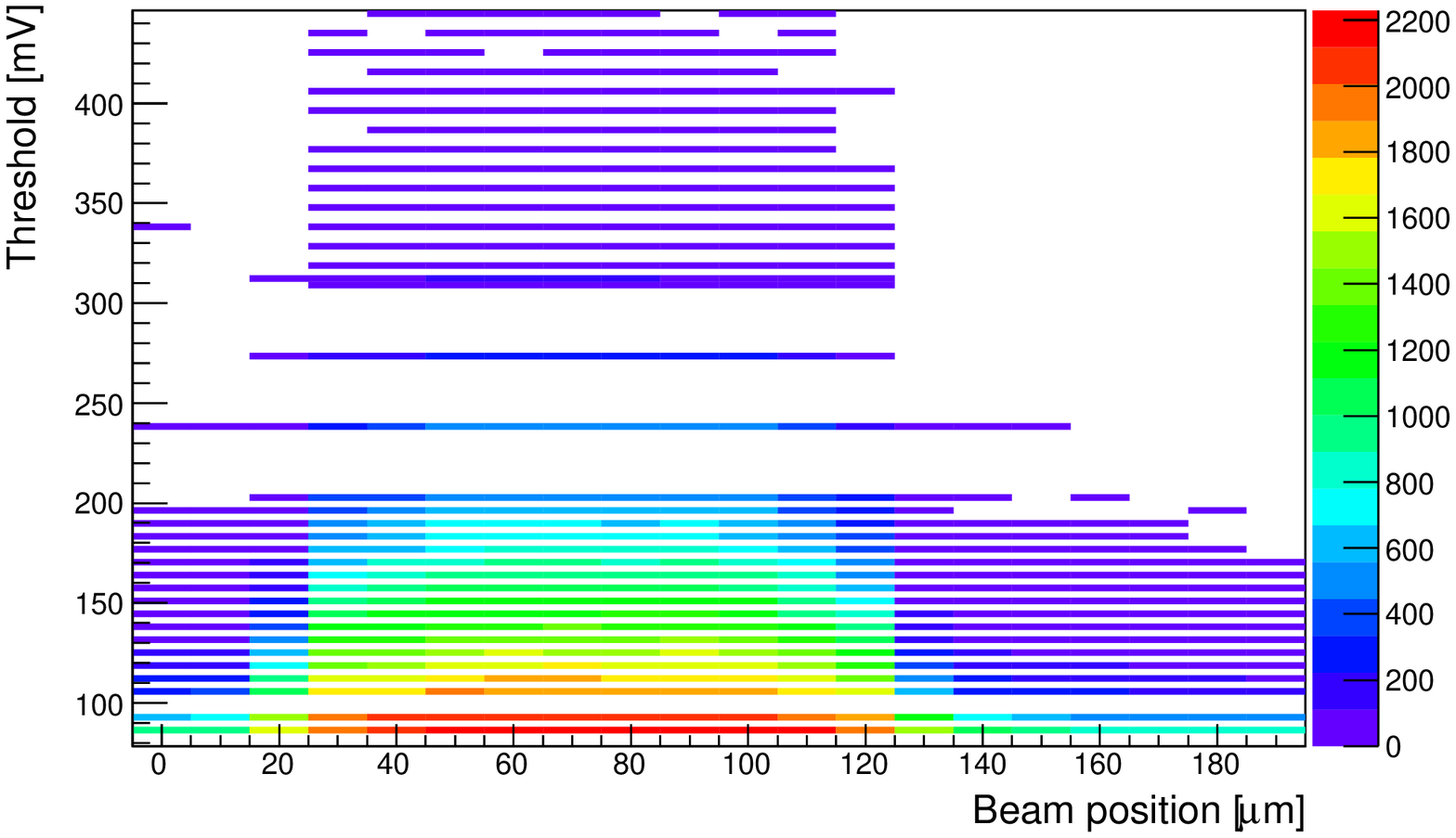}
  \caption{Hits obtained from one channel of an ABCN-25 module while scanning across the connected silicon sensor strip in steps of {\unit[10]{$\upmu$m}}.}
  \label{fig:6}
\end{subfigure}
\begin{subfigure}{.01\textwidth}
\hfill
\end{subfigure}%
\begin{subfigure}{.49\textwidth}
  \centering
  \includegraphics[width=\linewidth]{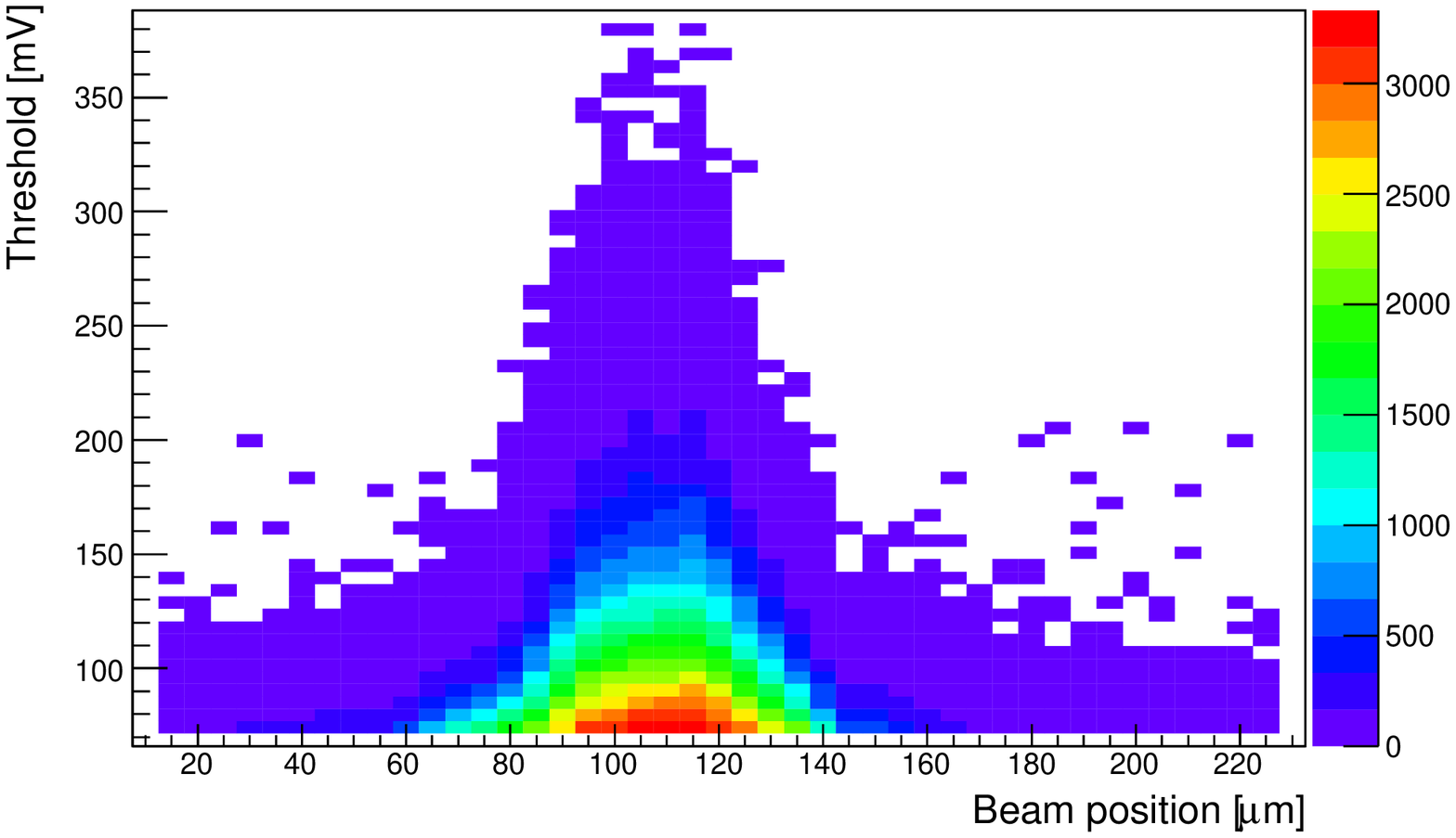}
  \caption{Hits obtained from one channel of an ABC130 mini module while scanning across the connected silicon sensor strip in steps of {\unit[5]{$\upmu$m}}.}
  \label{fig:11}
\end{subfigure}
\caption{Fine strip scans across three strips of a sensor for only the central strip, measured repeatedly at each beam position by using the same number of triggers for different readout thresholds. Due to time constraints, the scanning time for the ABCN-25 module (left) had to be reduced by increasing the position step size and decreasing the numbers of thresholds and triggers in the scan, leading to an overall lower number of hits per bin.}
\label{fig:611}
\end{figure}

Figure~\ref{fig:611} shows the results for threshold scans across three strips (\unit[10]{$\upmu$m} steps) of an end-cap sensor (see figure~\ref{fig:6}) and a barrel mini sensor (figure~\ref{fig:11}).
Due to different scanning parameters (see table~\ref{tab:scanpar}), in particular number of triggers and trimming, the numbers of obtained hits per position and thresholds differ between the two devices and do not allow a direct comparison of both devices. They do, however, allow for a comparison of the width of a sensor area connected to one readout channel with the nominal strip widths of each sensor.

For the ABCN-25 sensor, the hits obtained from an individual channel show the expected behaviour: over a scan distance of \unit[200]{$\upmu$m}, matching the sensor strip pitch in that region of \unit[103]{$\upmu$m}, the channel shows hits up to high thresholds. On the ABC130 module, the scan shows a detection range smaller than the expected strip pitch of \unit[74.5]{$\upmu$m} for the central strip (see below). Outside the region of the central strip, high rates were measured for low thresholds, corresponding to noise.

\subsection{Effective strip width}
\label{ss:esw}

\begin{figure}
\centering
\begin{subfigure}{.49\textwidth}
  \centering
  \includegraphics[width=\linewidth]{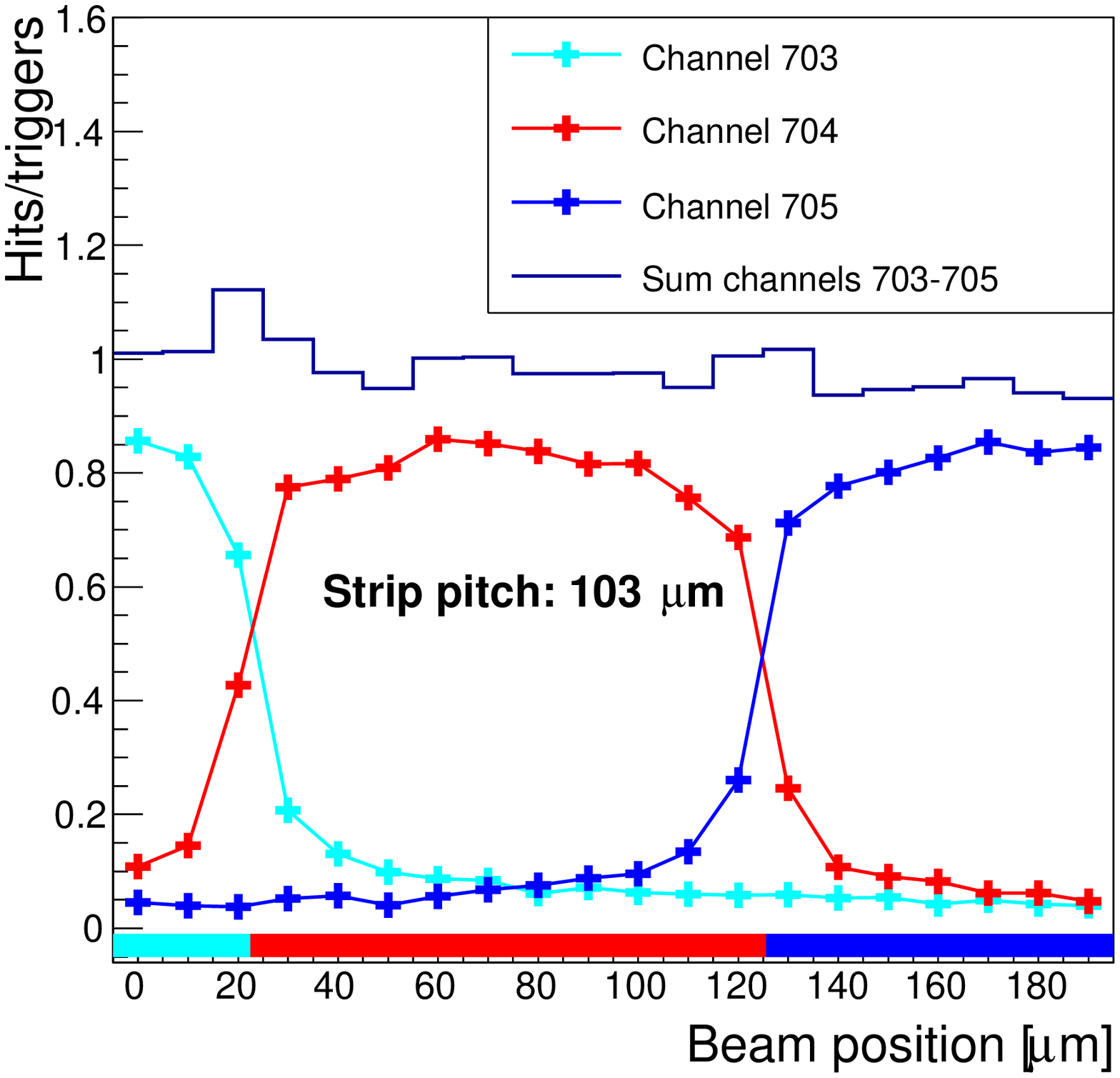}
  \caption{ABCN-25 module sensor scan: individual and combined hits from three adjacent strips for a fine strip scan across three strips (cyan, red and blue), showing the individual hits at a threshold of {\unit[105.6]{mV}}. The collected hits for the three channels fit both the positions of the strips and the sensor strip pitch well. This sensor does not have p-stop implants.}
  \label{fig:7}
\end{subfigure}
\begin{subfigure}{.01\textwidth}
\hfill
\end{subfigure}%
\begin{subfigure}{.49\textwidth}
  \centering
  \includegraphics[width=\linewidth]{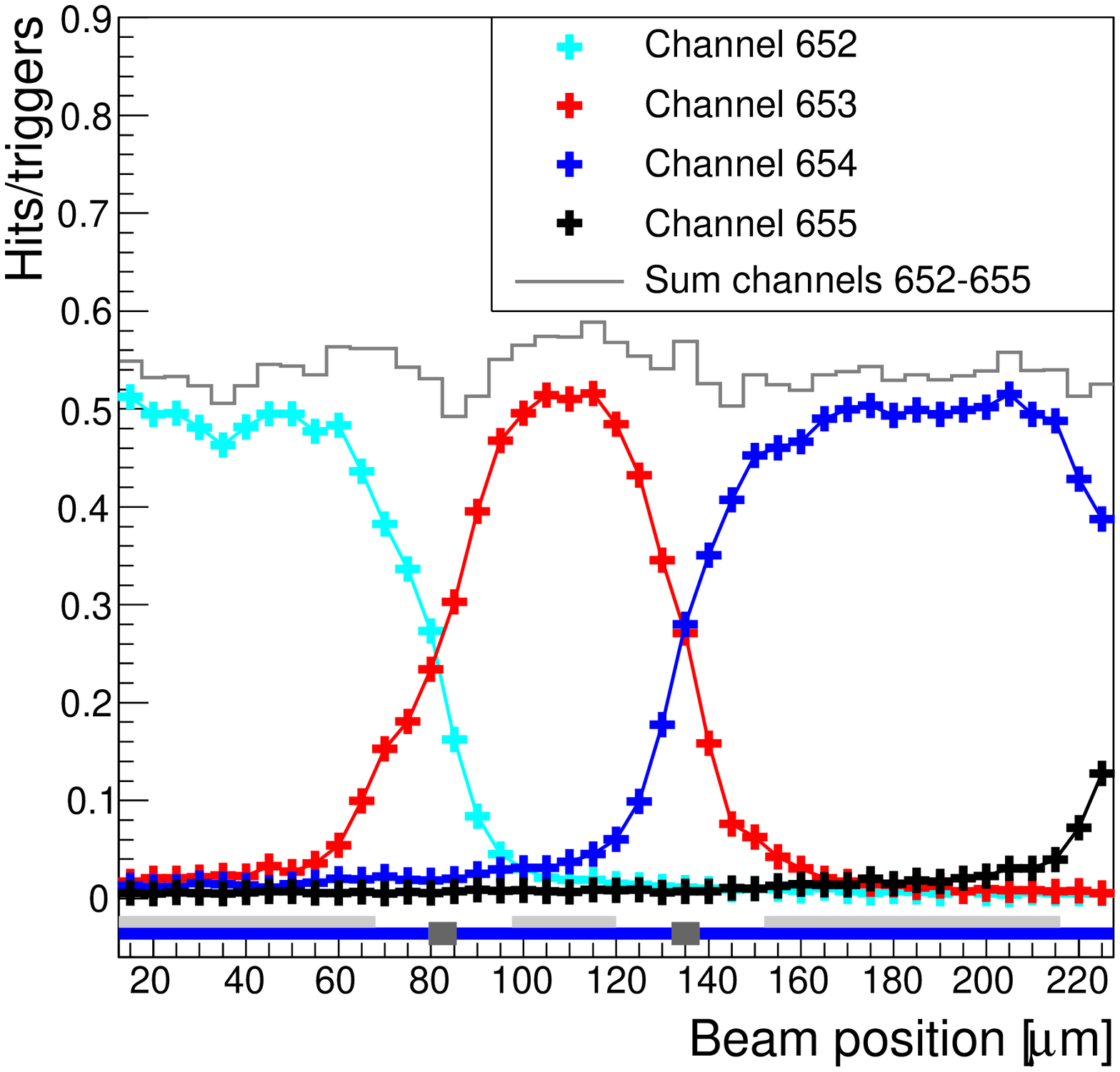}
  \caption{ABC130 module sensor scan: individual and combined hits from four adjacent channels at a possible operating threshold of {\unit[80]{mV}} with the corresponding sensor geometry: silicon (blue), aluminium layer (light grey) and p-stops (dark grey). The hits can be seen to agree well with the expected locations of p-stops and sensor strips.}
  \label{fig:10}
\end{subfigure}
\caption{Hit rates from several adjacent channels for an ABCN-25 module and an ABC130 mini module at given thresholds. The hit/trigger levels of both devices can not be compared directly due to different trimming settings and a reduced number of triggers for the ABCN-25 module (left), leading to its overall lower hit rate.}
\end{figure}

In order to relate the hits measured across individual strips to the sensor geometry, a reasonable threshold was selected to compare the hits from all channels connected to sensor strips covered in the scans. Figures~\ref{fig:7} and~\ref{fig:10} show the resulting strip shapes for the ABCN-25 endcap module and the ABC130 mini module, respectively.
For the ABCN-25 end-cap module, where the scan was performed in a sensor area without bond pads, all three strips under investigation show the same hit rates and shape across the strip. The width of the detection range matches the strip pitch of the sensor, as expected.

For the ABC130 module, the adjacent channels registered hits over areas of different widths. Comparing the width of the channels' detection ranges to the sensor geometry, the detected hits were found to agree well with the p-stop geometry in the measured area, indicating that the sensor geometry affects the effective width of a sensor strip. Independent of the width of the detection range, the hit rate at the central region of a channel is similar for all channels.

\section{Conclusion and outlook}

X-ray beam scans in steps of \unit[5]{$\upmu$m} and \unit[10]{$\upmu$m} were performed for two silicon strip detector modules for the ATLAS Phase-II Upgrade. The results show that the effective width of silicon strip sensors is determined by the local sensor geometry (bond pads on top of strips and p-stops between strips) rather than the pitch of the strip implants.


Future plans for the investigations of silicon strip modules foresee a crosscheck of the results found using an X-ray beam by using an electron particle beam. Additionally, the measurements performed for non-irradiated sensors are planned to be repeated using hadron irradiated silicon sensors in order to analyse changes in signal collection and efficiency caused by radiation damages.

\section*{Acknowledgements}

We thank Diamond Light Source for access to beamline B16 (proposal number MT11639) that contributed to the results presented here. The authors would like to thank personnel of the B16 beam, especially Andy Malandain and Julien Marchal for providing advice, support and maintenance during the experiment.

\section*{References}

\bibliographystyle{unsrt}
\bibliography{bibliography.bib}

\end{document}